# THE STATISTICAL ANALYSIS OF THE LIVE TV BIT RATE

Iskandar Aripov

Department of Computer Science, Georgia Institute of Technology, Atlanta, USA
iaripov3@gatech.edu

## ABSTRACT

*This paper studies the statistical nature of TV channels streaming variable bit rate distribution and allocation. The goal of the paper is to derive the best-fit rate distribution to describe TV streaming bandwidth allocation, which can reveal traffic demands of users. Our analysis uses multiplexers channel bandwidth allocation (PID) data of 13 TV live channels. We apply 17 continuous and 3 discrete distributions to determine the best-fit distribution function for each individual channel and for the whole set of channels. We found that the generalized extreme distribution fitting most of our channels most precisely according to the Bayesian information criterion. By the same criterion tlocationscale distribution matches best for the whole system. We use these results to propose parameters for streaming server queuing model. Results are useful for streaming servers scheduling policy design process targeting to improve limited infrastructural resources, traffic engineering through dynamic routing at CDN, SDN.*

## KEYWORDS

*Statistical Multiplexing, TV channels bandwidth utilization, Video compression & Streaming, Internet Signal Processing, queuing theory*

## 1. INTRODUCTION

The live TV streaming services over the Internet, air, satellite, mobile or even cable, are getting popular day by day. To achieve bandwidth efficient allocation through diverse media it is essential to control the streaming speed to use resources appropriately. Therefore the TV streaming bit rate nature identification and respective network simulation play an important role. For these reasons we got interested to study the nature of the TV stream considering it in SD streaming quality. To try to maximize the efficiency of the network bandwidth usage we decided to investigate short sampling statistics of available data. We processed it using MATLAB and a supplementary script [1]. The streaming data was obtained from ReflexTM Statistical Multiplexing by Ericsson, which architecture is described in [2] and configuration in [3]. Multiplexer is responsible for sampling of 13 live TV channels at a major TV Streaming Station of one of the CIS countries on Head-End side.

We want to derive statistical distribution which best approximates the bandwidth allocation of the system. In doing so we are able to describe the traffic demands of the data streams. There had been plenty of modeling work in Statistical Multiplexing such as [4], [5], [6] of Multiple time scale Markov streams, integrated service networks, and H.264 Video Streams Using Structural Similarity Information respectively. [7] focuses on describing how the architecture of the system can be modified to take advantage of statistical multiplexing for DVB-H channels which are designated for mobile users. There are also plenty of studies on algorithmic and architectural enhancement of video codec design such as [8], but a few available studies that take real world TV samples and attempts to extract the statistical nature of multiplexers bandwidth allocation. We discuss more papers in section V Relevant work. Merging statistical approaches with real world data can help us to develop recommendations on bandwidth allocation and multiplexer design by merging existing theory with the real world data.

To determine the distribution we chop the data into histogram bins and fit 17 continuous and 3 discrete distributions to determine the best-fit distribution function for each individual channel and for the whole set of channels based on Bayesian information criterion. Then we determine whether multiplexer channel allocation follows any statistical law and based on that observation we propose recommendations in multiplexer design. Our paper is organized as follows: Section 2 presents Measurement Framework describing in detail about the data, parameters, and how we carry out experiment. Section 3 presents Results and interpretation of the experiment. Section 4 describes parameters for queuing model based on results of Section 3. Section 5 talks more about relevant work and Section 6 concludes the paper.

## 2. MEASUREMENT FRAMEWORK

We will refer to channel bandwidth allocation as a PID. PID stands for proportional-integral-derivative controller and in our framework it is how much bandwidth is allocated by a multiplexer for a channel at a time sample was taken. We obtained 23 minutes of channel bandwidth allocation (PID) data in Kbps (so value 2000 corresponds to 2Mbps), for a total of 13 channels and 36673 measurements (2821 per channel). The channel categories include: National News, City News, Sport, Movies, Cultural, Music, Kids, Family, Community, Show business. Different channels have a different bandwidth allocated. This can be explained by a priority, for example breaking news or live sporting events will require higher bandwidth at the expense of less urgent channels. Most low priority channels are allocated 2.5 or 2 Mbts, some higher priority channels can get 5Mbts and the bandwidth for the highest priority channels can exceed 6 Mbts. More details are available in results section. The multiplexer measurements took place between 18:39 and 19:02 of local time when many people of that country usually gather to watch TV. Our measurements include two sets: selected channel and the overall system. First we draw the channel bandwidth to samples graph to give intuition on bandwidth distribution. Then we determine the best-fit distribution, we chop the data into histogram bins and fit 17 continuous and 3 discrete distributions to determine the best-fit distribution function for each individual channel and for the whole set of channels based on Bayesian information criterion. We compute the histogram with 4 best fitting curves for selected channel and overall system. Then we derive a table that describes what distribution is 1st, 2nd, or 3rd best fit to all channels and to how many channels.

Note: The experimental environment consists of Matlab, script, and data set.

## 3. RESULTS

### 3.1 Selected single channel

Table 1 and 2 contain the following characteristics:

BIC - Bayesian information criterion which is used in statistics as a model selection based on value of the maximum likelihood function where higher maximum likelihood function means better fit.

PDF – Probability density function of the distribution.

[k, σ, μ, ν] – parameters specific to the pdf of each distribution that given numerical value can recreate that best-fit distribution.

Table 1: Best-fit distributions parameters for selected culture channel

| Distribution Name | BIC | PDF f(x) | [k-shape, σ-scale, μ-location, ν-shape] |
|---|---|---|---|
| tlocationscale | 44010.07 | $\dfrac{\Gamma\left(\frac{\nu+1}{2}\right)}{\sigma\sqrt{\nu\pi}\,\Gamma\left(\frac{\nu}{2}\right)}\left[\dfrac{\nu+\left(\frac{x-\mu}{\sigma}\right)^2}{\nu}\right]^{-\left(\frac{\nu+1}{2}\right)}$<br>$\Gamma(*)$ - gamma function | $[\mu, \sigma, \nu] =$ [1644.76, 454.92, 4.137] |
| logistic | 44023.37 | $\dfrac{\exp\left\{\frac{x-\mu}{\sigma}\right\}}{\sigma\left(1+\exp\left\{\frac{x-\mu}{\sigma}\right\}\right)^2}$;<br>$-\infty < x < \infty$ | $[\mu, \sigma] =$ [1660.73, 325.24] |
| normal | 44195.08 | $\dfrac{1}{\sigma\sqrt{2\pi}}e^{\frac{-(x-\mu)^2}{2\sigma^2}}, for\ x \in R$ | $[\mu, \sigma] =$ [1696.09, 607.21] |
| generalized extreme value | 44209.92 | $\left(\dfrac{1}{\sigma}\right)\exp\left(-\left(1+k\dfrac{(x-\mu)}{\sigma}\right)^{-\frac{1}{k}}\right)\left(\left(1+k\dfrac{(x-\mu)}{\sigma}\right)^{-1-\frac{1}{k}}\right)$ | $[k, \sigma, \mu] =$ [-0.19, 587.27, 1460.07] |

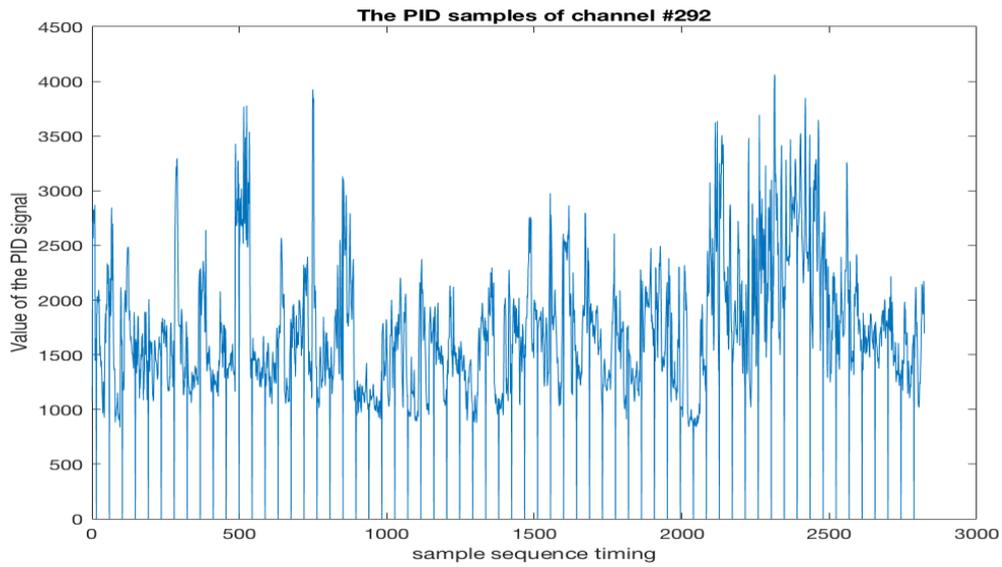

Figure 1: The bandwidth allocated by the multiplexer for the selected culture channel during 2821 samples and 23 minutes. Y-axis refers to bandwidth and is in Kbps

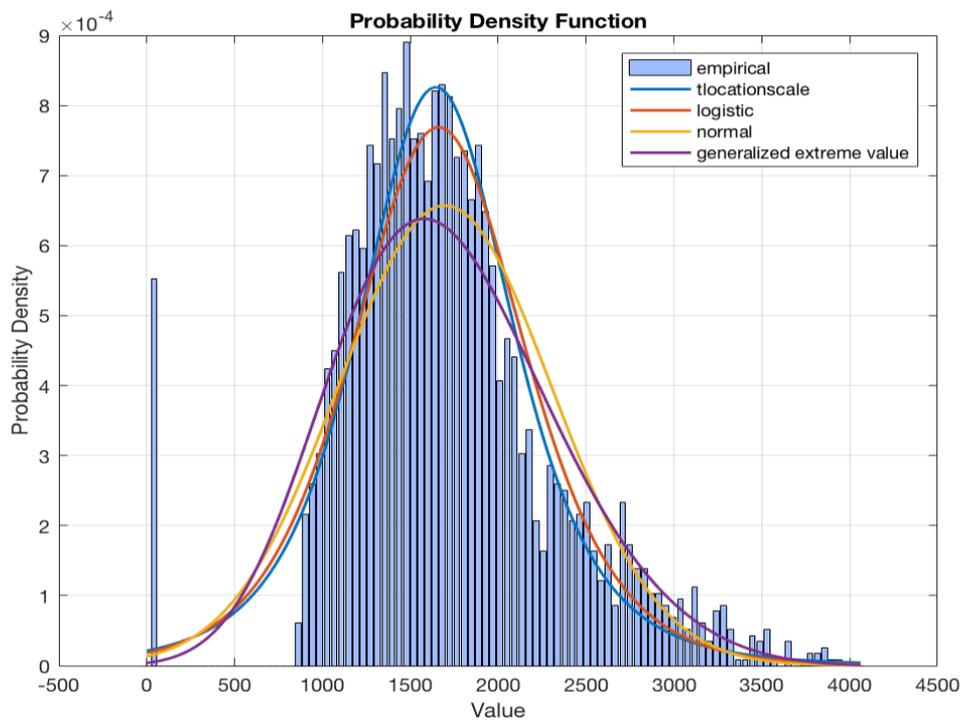

Figure 2: The Probability density function of the histogram of the selected channel and best-fit distribution fitting. X-axis refers to bandwidth and is in Kbps. We see peak pdf values correspond to top to bottom order of distributions listed by Matlab from greatest to smallest values respectively.

## 3.2 Results for all channels

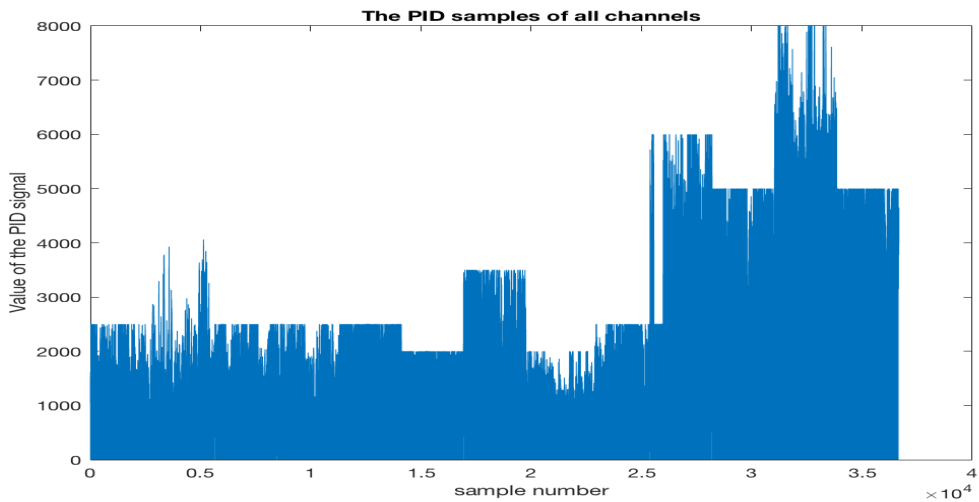

Figure 3: The bandwidth allocated by the multiplexer to every channel. Y-axis refers to bandwidth and is in Kbps.

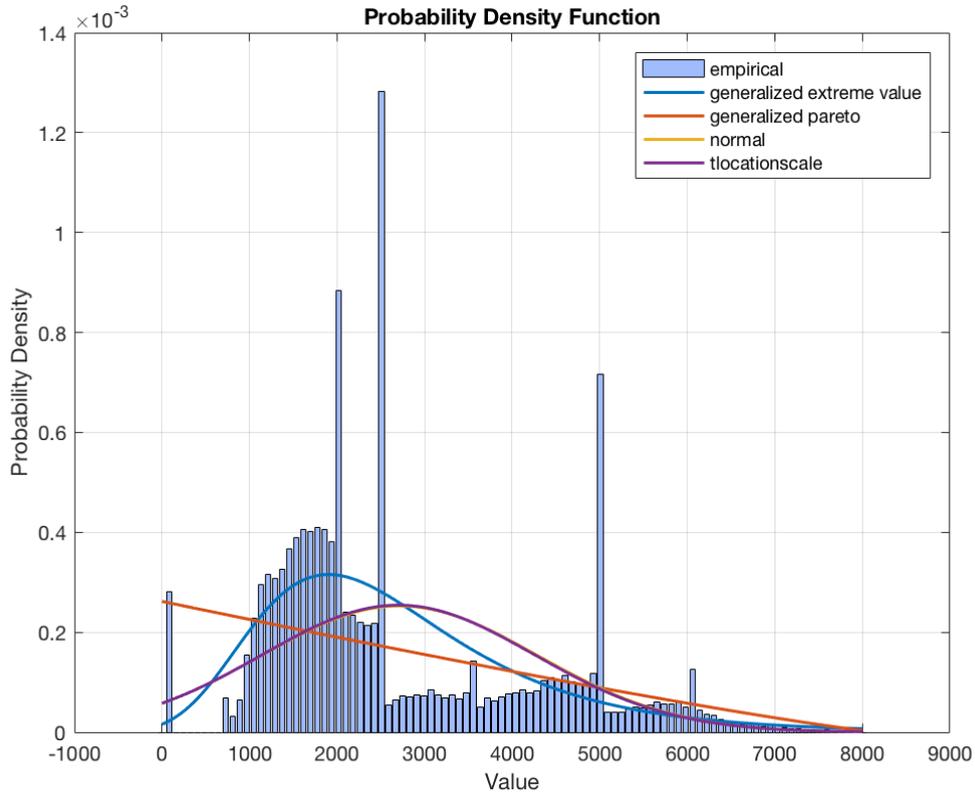

Figure 4: The PDF of the histogram for all the channels and best-fit distribution fitting. X-axis refers to bandwidth and is in Kbps. We see peak pdf values correspond to top to bottom order of distributions listed by Matlab from greatest to smallest values respectively excluding normal distribution.

Table 2: Best fit distributions parameters for all channels histogram.

| Distribution Name | BIC | PDF f(x) | [k-shape, σ-scale, μ-location, v-shape] |
|---|---|---|---|
| generalized extreme value | 636526.71 | $\left(\frac{1}{\sigma}\right) exp\left(-\left(1+k\frac{(x-\mu)}{\sigma}\right)^{-\frac{1}{k}}\right)\left(\left(1+k\frac{(x-\mu)}{\sigma}\right)^{-1-\frac{1}{k}}\right)$ | $[k, \sigma, \mu] =$ [0.06, 1167.89, 1965.89] |
| generalized pareto | 644104.36 | $\left(\frac{1}{\sigma}\right)\left(1+k\frac{(x-\mu)}{\sigma}\right)^{-1-\frac{1}{k}}$, for $\theta < x$, when $k > 0$, or for $\theta < x < \theta - \sigma/k$ when $k < 0$. | $[k, \sigma, \theta] =$ [-0.47, 3820.24, -2.22e-15] |
| normal | 644279.58 | $\frac{1}{\sigma\sqrt{2\pi}} e^{\frac{-(x-\mu)^2}{2\sigma^2}}, for\ x \in R$ | $[\mu, \sigma] =$ [2712.56, 1579.56] |
| tlocationscale | 644287.53 | $\frac{\Gamma\left(\frac{v+1}{2}\right)}{\sigma\sqrt{v\pi}\Gamma\left(\frac{v}{2}\right)}\left[\frac{v+\left(\frac{x-\mu}{\sigma}\right)^2}{v}\right]^{-\left(\frac{v+1}{2}\right)}$ $\Gamma(*)$ - gamma function | $[\mu, \sigma, v] =$ [2701.14, 1566.54, 121.73] |

Table 3: For how many channels does each distribution 1st, 2nd, or 3rd best-fit. Overall refers to all channels combined.

| Distribution Name | 1st best fit relative # of channels | 2nd best fit relative # of channels | 3rd best fit relative # of channels |
|---|---|---|---|
| tlocationscale | 3 | 2 | |
| logistic | | 2 | 1 |
| normal | | | 1, overall |
| generalized extreme value | Overall system of 13 channels | 8 | 2 |
| extreme value | | | 9 |
| generalized pareto | 10 | 1, overall | |
| exponential | | | |

Selected channel from cultural category from Figure 1 was chosen for demonstration because of its signal non-uniformity relatively to other channels. Such non-uniformities are because of the content itself. Certain video content, such as movies with a lot of action or sporting events have higher traffic requirements than for example news reporting where there are minimal changes between frames. Given this information we can presume that selected channel was having change of scene multiple times. Perhaps a concert was played, given that it is from cultural category. We can say that selected channel is of the lower priority based on the fact that it was allocated 2.5Mbits. Based on Figure 2, its probability density function happened to demonstrate most resemblance to tlocationscale distribution. Selected channel can be found as 2nd channel in the overall graph of all channels by it signal shape graph. Figure 3 shows us bandwidth allocated to all 13 channels. We can see some contrast of channels, which require higher bandwidth on average than others. This is attributed by the higher priority and quality of streaming these channels require. We also see non-uniformities within the channels such as in selected channel.

In Figure 4 we examine the probability density function of all 13 channels. We can see that if we asked to model the average channel capacity for 13 given, then our most accurate choice will be "generalized extreme value" distribution with parameters given in Table 2. However, if we are asked to model one random channel, then according to the Table 3 above we will pick "generalized pareto distribution".

To verify our analysis of signal non-uniformity we computed Hurst exponent (using [9]) for all channels and all 13 channels had exponents in 0-0.4 range. This means that channel bandwidth allocation is non-random and follows mean-reverting series: an increase is likely followed by decrease or vice versa with tendency to return to long-term mean. These results explain non-uniformities from section III and are to be expected because multiplexer might temporary allocate higher bandwidth for some high intensity video segment, followed by lowering bandwidth for selected channel to revolve around its allocated bandwidth.

To model the synchronized bandwidth requirement for the whole multiplexor with 13 channels we need to synchronize our channels bandwidth consumption (stack up values of bandwidth allocation for every channel at the same time). Then we will be able to see how much bandwidth each channel requires at a certain time. Figure 5 is the graph of all 13 channels per corresponding samples.

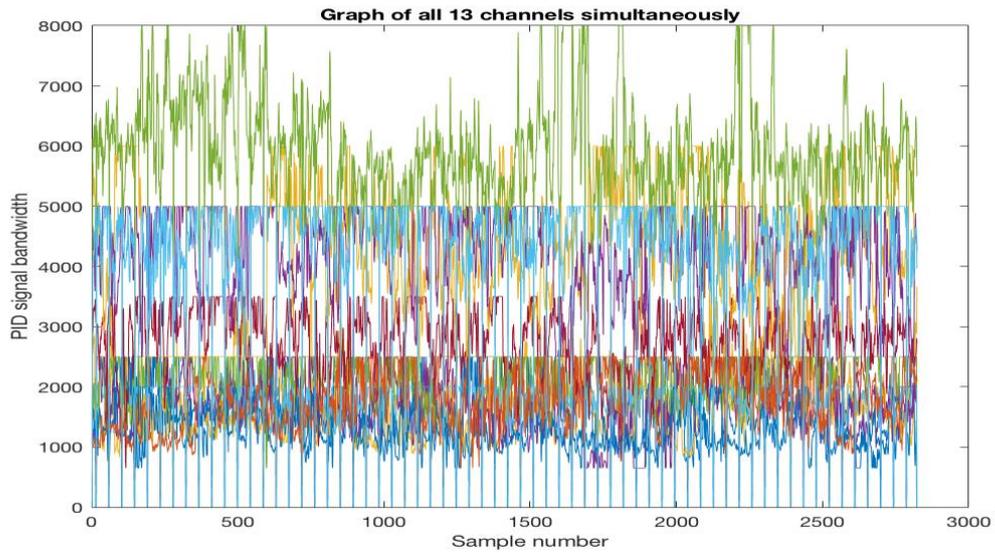

Figure 5: All channels synchronized bandwidth allocation

After we synchronized our data, we need to add up the bandwidth consumption of all channels at a particular sample. In Figure 6 we get a uniform graph between capacities of 35-40 Mbts, which is consistent with allocated bandwidth.

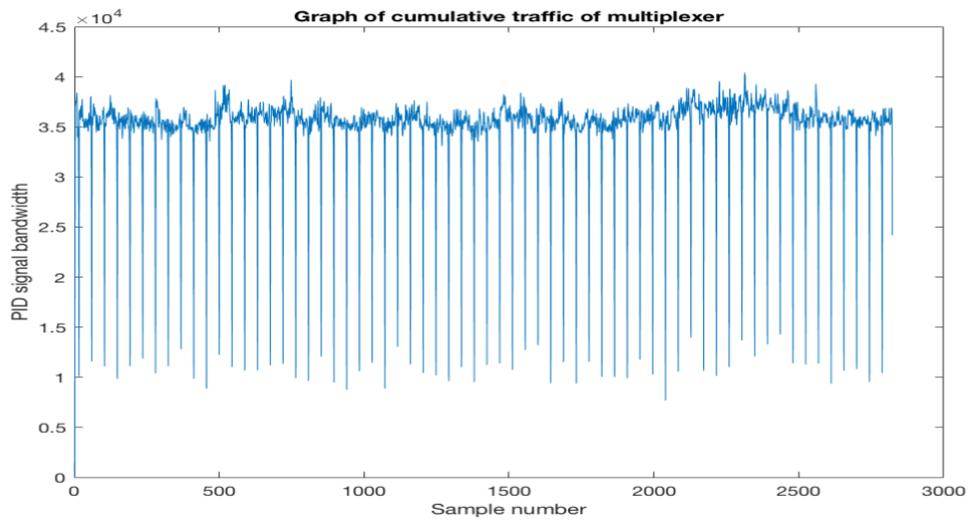

Figure 6: The bandwidth allocated by the multiplexer to 13 channels synchronized during 23 minutes. The Y-axis is in Kbps.

Now computing the histogram of our multiplexor in Figure 7.

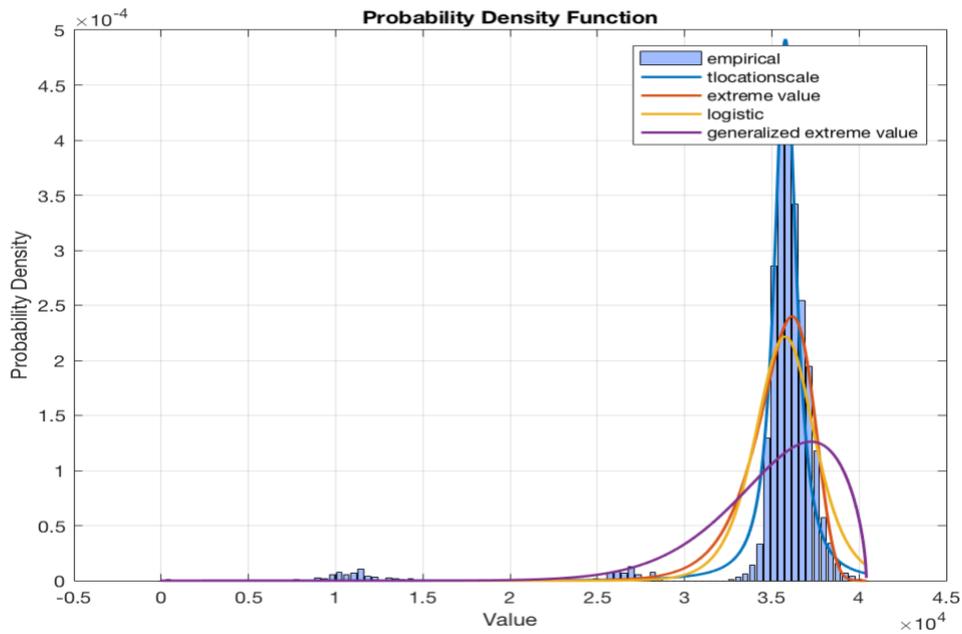

Figure 7: The PDF of the multiplexer's bandwidth allocation. The X-axis is in Kbps. We see peak pdf values correspond to top to bottom order of distributions listed by Matlab from greatest to smallest values respectively.

We see a very close resemblance to the tlocationscale distribution. Indeed, multiplexer bandwidth allocation is statistical in nature for this dataset.

## 4. Follow-up Analysis

After we developed statistical models for individual channels and Multiplexer, we can determine queuing model needed to simulate behaviour of streaming server and calculate its capacity depending on traffic loads. It can be used to compute Streaming server bandwidth needed as a function of number of users. We model streaming server as a buffer, which receives traffic through CDN from multiplexer. Figure 8.1 and 8.2 demonstrate input to streaming server as an output of the multiplexer. Figure 8.2 demonstrate output of streaming server as a cumulative sum of traffic consumed by groups of users.

We use the following formula for traffic consumed by each group:

Streaming Server bandwidth = $\sum$ Group M total traffic

= $\sum$ N * PID(M) for all M.

N = number of users watching channel M

PID(M) = Bandwidth allocated to channel M by multiplexer N can be approximated by Poisson distribution with expectation = $\lambda$. PID(M) each can be approximated with best fit distributions specific to that channel. For example, PID(1) which is culture channel can be approximated with "tlocationscale" distribution with parameters from table 1. As we determined in Table 3, we can model most PID channels with "generalized pareto" distribution.

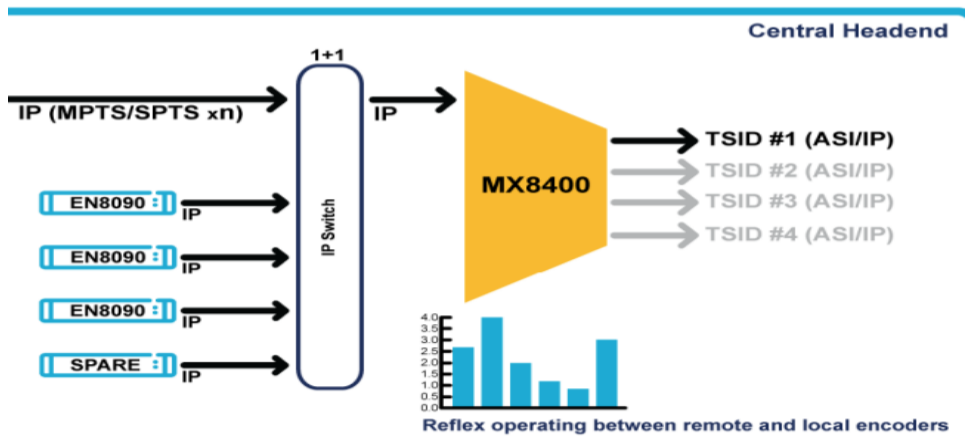

Figure 8.1 [2]: Multiplexer controls channels bandwidth allocation

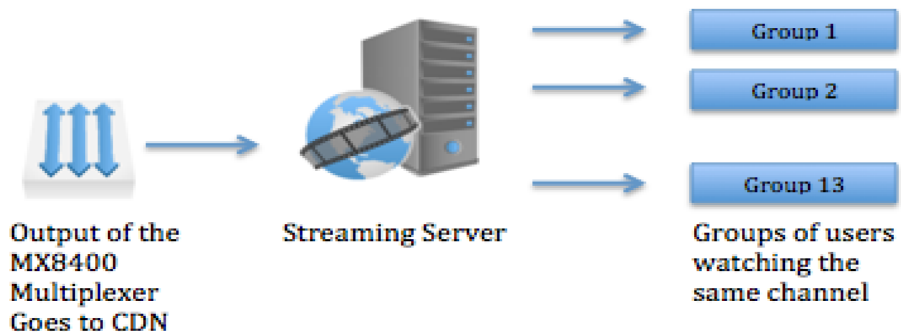

Figure 8.2: Streaming server shows channels to different streaming groups of users- watching the same channel.

## 5. Related Work

The work of [10] demonstrated that the network traffic shows self-similar pattern. We verified their results specifically for live streaming multimedia.

Other relevant papers [11], [12] do similar traffic analysis but with the target to specifically underlying P2P architecture using famous PPLIVE video streaming platform. Such architectures are effective only if there is sufficient number of peer users uploading and downloading content at the same time and who have sufficient uploading capacity. Our system is different because we assume standard client-server model to distribute content to users. But similarly to our study of bandwidth vs sample number, [11], [12] study dynamics of Number of peers vs time. Other important study is [13] with very similar system for DVB steam analysis from which we gathered our dataset. The authors talk about different modes of Statistical multiplexing and go in detail of the architecture of Statistical Multiplexer and of advantages of statistical multiplexing in DVB TV transmission. But they don't use samples to model multiplexer distribution, rather they provide broader overview of statistical multiplexing for TV transmission. Usually streaming servers utilize M/M/k or M/G/k queuing models to determine needed capacity. Our calculations make more efficient use of bandwidth by catering the queuing model to specific statistical distributions of the live TV traffic. [14] verified DVB compliance of statistical multiplexing for digital terrestrial television of 10 Colombian channels over 100 seconds.

Other more recent works focused on more specific system subparts for bandwidth preservation. [15] described perceptually inspired video processing technology which improves video compression for modern standards. [16] proposed using frame frequency error optimization technique based on Ultra-high efficiency video coding. [17] proposed novel cross-layer packet prioritization scheme for error-resilient transmission. [18] suggested activating the interaction between Set Top Box (STBs) and unicast streaming for faster channel change time. These bandwidth utilization improvements are applicable in conjunction with our findings.

## 6. Conclusion and Future direction

Our analysis models the bit rate for individual and for all channels combined and our results suggest that for most individual channels bandwidth is allocated with "generalized extreme value" and for all channels combined "generalized pareto" distribution is appropriate. We also simulated the bandwidth allocation of the multiplexer and got very accurate resemblance with "tlocationscale" distribution. Such analysis helped us to better understand the statistical nature of the TV traffic allocation. Obtained results are important for designing of CDN streaming server queuing model. The results can be used for wide range of applications and in particular to simulate the behaviour of rate-based or/and buffer-based servicing discipline of adaptive bit rate at streaming clouds set up.

### Acknowledgement

We would like to thank Professor Klara Nahrstedt, who provided feedback and expertise that greatly assisted this research paper.

**Author**

Iskandar Aripov is a Master of Science student in computer science at Georgia Institute of Technology with a focus on computing systems. He holds a Bachelor of Science degree in electrical engineering from University of Illinois at Urbana-Champaign. His research interests are in computer networks, multimedia systems, and cyber security.